\documentstyle[aps,prl,epsf,rotate]{revtex}

\begin{document}

\twocolumn[\hsize\textwidth\columnwidth\hsize\csname
@twocolumnfalse\endcsname

\title{Resistance spikes and domain wall loops in Ising quantum Hall ferromagnets}

\author{T. Jungwirth$^{1,2}$ and A.H. MacDonald$^{1}$ 
}
\address{
$^{1}$Department of Physics, The University of Texas, Austin, TX 78712}
\address{
$^{2}$Institute of Physics ASCR, Cukrovarnick\'a 10,  162 53 Praha 6, Czech Republic
}

\date{\today}
\maketitle

\begin{abstract}
We explain the recent observation of resistance spikes and
hysteretic transport properties in Ising quantum Hall ferromagnets
in terms of the unique physics of their domain walls.
Self-consistent RPA/Hartree-Fock theory is applied to
microscopically determine properties of the ground state
and domain-wall excitations. 
In these systems domain wall loops  support
one-dimensional electron systems with an effective mass comparable
to the bare electron mass and may carry charge.
Our theory is able to account quantitatively for the experimental
Ising critical temperature and to explain  characteristics
of the resistive hysteresis loops.
\end{abstract}

\vskip2pc]
%

For two-dimensional electrons in a perpendicular magnetic field $B_{\perp}$,
independent electron eigenstates occur in manifolds known as Landau
levels with macroscopic degeneracy $AB_{\perp}/\Phi_0$, 
where $A$ is the sample area and $\Phi_0$ is the magnetic flux quantum. 
The zero-width energy bands are responsible for a tremendous
variety of many-body physics that has been observed in the quantum Hall
regime \cite{prangegirvin,sarmapinczuk}.
Quantum Hall ferromagnetism, of interest here, occurs when two different Landau
levels distinguished by the cyclotron energy, spin, or quantum well subband labels
of their orbitals are brought into energetic alignment and the 
Landau level filling factor $\nu$ is close to an integer.
Neglecting 
charge fluctuations, low-energy states of quantum Hall ferromagnets (QHFs) are 
specified 
by assigning 
to each orbital in the Landau level
a two-component spinor $(\cos\theta/2,
e^{i\varphi}\sin\theta/2)$ corresponding to a pseudospin oriented
along a general unit vector $\hat{m}=(\sin\theta\cos\varphi, \sin\theta\sin\varphi,
\cos\theta)$. (The influence of remote Landau levels can be captured 
perturbatively as necessary.)
While ordered states can occur when any
two Landau levels simultaneously approach the chemical
potential, the nature of the ground state is sensitive to the
microscopic character of the crossing Landau
levels \cite{jungwirthprl98,jungwirthprb01}.
Isotropic \cite{tycko,girvinmacd},
XY \cite{girvinmacd,eisenstein,spielman}
and
Ising QHFs
\cite{jungwirthprl98,piazza,eom,giuliani,daneshvar}
are now well established.
Our work is motivated by the recent observation \cite{depoortere}
of hysteretic transport and unexplained resistance spikes
when Landau levels with different quantized kinetic (cyclotron) energies cross.
We argue that the resistance spikes are due to charge transport in the
1D quasiparticle systems of long domain wall loops
and establish a correspondence between their occurrence
and vanishing domain-wall free-energy density at the Ising transition
temperature, $T_c$.

The dependence of the uniform QHF state energy per electron
on pseudospin orientation has the form \cite{jungwirthprb01}:
$E[\hat{m}]=-bm_z-Jm_z^2$, where $b$ is an effective magnetic field
that includes both single-particle Landau level splitting and interaction
contributions \cite{jungwirthprb01}
and $J >0$ is an effective 
Ising interaction parameter. 
At $b =0$, the $m_z = 1$ (pseudospin $\uparrow$) and 
$m_z = -1$ (pseudospin $\downarrow$) states are degenerate.
In the following we establish an association between $b=0$ 
and  
the experimental resistance spikes, and propose an explanation
for the spike origin.  
For tilted magnetic fields and variable 2D electron densities, the
$b=0$ condition at a given filling factor is achieved along 
a continuous line in the two-dimensional $(B_{tot}-B_{\perp})$ space, 
which can be explored experimentally by tilting the field away
from the sample normal.
($B_{tot}$ is the total magnetic field.) 
\begin{figure}[h]
\epsfxsize=3.3in
\hspace*{-0.5cm}
\centerline{
\epsffile{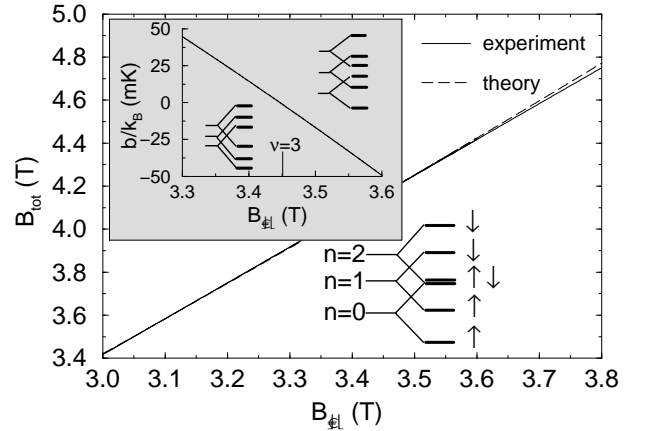}}
\caption{
Main graph: Theoretical (dashed) line
of the $n=2$ spin-up and $n=0$ spin-down level
coincidences in the perpendicular magnetic field --
total field space,
calculated using nominal sample parameters of De Poortere {\em et al.}
\protect\cite{depoortere}
and taking into account  RPA-screening of the Coulomb interaction.
The density  of the 2D electron system is
2.5~$\times 10^{11}$~cm$^{-2}$.
The solid line in the main graph
is a fit to the experimental magnetoresistance spike positions
presented in Ref.~\protect\cite{depoortere}
in Fig. 3(A).
Inset: 
Effective external field of the Ising model calculated as a function
of the perpendicular magnetic field at a fixed field tilt angle 34$^o$.
Landau level configurations at positive
and negative values of the $b$--field are also indicated in the inset.
}
\label{prlfig1}
\end{figure}
In Fig.~\ref{prlfig1}
we compare our theoretical\cite{jungwirthprl98,jungwirthprb99}
$b=0$ line for $\nu=3$, based on numerical
self-consistent-field calculations for the geometry of
De Poortere {\em et al.}'s sample and on
many-body RPA/Hartree-Fock theory, with the line along
which resistance spikes were observed.  The coincidence
of these two curves strongly suggests that the spikes occur when   
$b=0$.  The same calculations\cite{jungwirthprb01} yield the
estimate $J= 0.018 e^2/\epsilon \ell/k_B \sim
2  {\rm K}$, where $\ell$ is the magnetic
length defined by $2 \pi \ell^2 B_{\perp} = \Phi_0$.

The ground state of an Ising QHF has $m_z = 1$
for $b>0$ and $m_z = -1$ for $b <0$.
At finite temperatures, non-trivial pseudospin magnetization configurations
become important.  For Ising QHFs, an elementary calculation
shows that spin-wave collective excitations have a gap $\Delta/k_B  = 4J/k_B
\sim 8 {\rm K}$.  Since the hysteretic resistance
spikes occur only for $T < 0.5 {\rm K}$,  spin-wave
excitations cannot play a role.
Instead, as we now
explain, the important thermal fluctuations in Ising QHFs
involve domain walls between $m_z=1$ and $m_z=-1$ regions
of the sample.

In classical 2D Ising models the critical temperature can be  
understood as a competition between unfavorable near-neighbor-spin interaction 
energy
along a domain wall, $L\gamma$, and the wall configurational 
entropy, 
$Ls_c=L/\xi\, k_B\ln (3)$, where $\xi$ is the domain wall
persistence length.  Both  
give free-energy contributions
proportional to wall length, $L$, with the former effect favoring short walls
and the latter contribution, which is proportional to temperature,
favoring long walls.  For $T > T_c$ the system 
free energy is lowered when domain walls expand to the sample perimeters,
destroying magnetic order.   The structure of domain 
walls is more complicated in Ising QHFs.  
As the domain wall is transversed,
the local pseudospin orientation goes from the north 
pole ($m_z=1$) to the south pole ($m_z=-1$), at a fixed orientation $\varphi$ of 
its $\hat x-\hat y$ plane projection.
We have evaluated the energy per unit length $\gamma$ of an infinite domain wall
by solving self-consistent Hartree-Fock equations:
\begin{equation}
\tan\theta(X)=\frac{-2\big(b-H^F_{\uparrow,\downarrow}(X)\big)}
{H^H_{\uparrow,\uparrow}(X)+H^F_{\uparrow,\uparrow}(X)
-H^H_{\downarrow,\downarrow}(X)-H^F_{\downarrow,\downarrow}(X)}\; ,
\label{hf}
\end{equation}
where the Hartree energy is given by
\begin{eqnarray}
& &H^H_{\sigma,\sigma^{\prime}}(X)=\sum_{X^{\prime}}\int d^3\vec{r}_1
\int d^3\vec{r}_2 V(\vec{r}_1-\vec{r}_2)\,\times\nonumber \\
& &\psi^{\ast}_{\sigma,X}(\vec{r}_1)
\psi_{\sigma^{\prime},X}(\vec{r}_1)
\psi^{\ast}_{\hat{m}(X^{\prime}),X^{\prime}}(\vec{r}_2)
\psi_{\hat{m}(X^{\prime}),X^{\prime}}(\vec{r}_2)
\label{hartree}
\end{eqnarray}
and the exchange energy by
\begin{eqnarray}
& &H^F_{\sigma,\sigma^{\prime}}(X)=-\sum_{X^{\prime}}\int d^3\vec{r}_1
\int d^3\vec{r}_2 V(\vec{r}_1-\vec{r}_2)\,\times\nonumber \\
& &\psi^{\ast}_{\sigma,X}(\vec{r}_1)
\psi_{\sigma^{\prime},X}(\vec{r}_2)
\psi^{\ast}_{\hat{m}(X^{\prime}),X^{\prime}}(\vec{r}_1)
\psi_{\hat{m}(X^{\prime}),X^{\prime}}(\vec{r}_2)\; .  
\label{exchange}
\end{eqnarray}
In Eqs.~(\ref{hartree}) and (\ref{exchange}), $V(\vec{r}_1-\vec{r}_2)$ is
the RPA-screened Coulomb potential and 
the self-consistent-field one-particle orbitals,
$\psi_{\sigma,X}(\vec{r})$,
are extended along the domain wall 
and localized 
near wavevector $k$ dependent guiding centers $X=k \ell^2$.
The energy density $\gamma$ is proportional to the increase 
in Hartree-Fock quasiparticle energies, 
integrated across the domain wall.
We find that the domain wall width is typically several magnetic lengths
and for the $\nu=3$ coincidence we find that $\gamma \ell = 
0.009 e^2/\epsilon \ell$.

A unique property of QHFs is the proportionality 
between electron charge density and pseudospin topological index 
density. It is this property that is responsible for the 
fascinating skyrmion physics
extensively studied in the isotropic case \cite{tycko,girvinmacd}.
In the case of Ising QHFs, the proportionality
implies a local excess charge per unit length along a domain wall 
$\rho_{\parallel} = e \nabla \varphi \cdot \hat n/(2 \pi) $ 
where $\hat n$ specifies
the local direction along the domain wall.  Single-valuedness of the magnetization
requires that the winding number of the angle $\varphi$ around a 
domain wall loop be quantized in units of $2 \pi$ and hence that 
the excess charge of a domain wall loop be quantized in units of the the 
electron charge $e$. The free-energy associated with the 
classical $\varphi$ field fluctuations within a domain wall,
\begin{equation}
f_{\varphi}=\frac{1}{L}k_BT\ln Z ;\;
Z=\int{\cal D}\varphi\exp\big(E_c[\varphi]/k_BT\big)\; ,
\label{phifree}
\end{equation} 
is controlled by the Coulomb interaction energy $E_c$
due to the consequent charge
fluctuations.

Assuming a domain wall persistence length $\xi\approx\ell$,
the free energy density of Ising QHF
domain wall
loops is given by
$f=k_BT\ln(3)/\ell+\gamma+f_{\varphi}$ and equals zero at $T=T_c$.
For the $\nu=3$ QHF, these considerations imply
that infinitely long domain walls proliferate and order is lost
for $T$ larger than the transition temperature
$T_c\approx 500$~mK.  
The close correspondence between this $T_c$ estimate, and 
the maximum temperature ($430$~mK) at which hysteretic resistance spikes  
are observed\cite{depoortere}
strongly supports 
our contention that the unusual transport phenomena are a 
consequence of the existence of long domain wall loops in these 
materials.  
In the following we  
first discuss the $b$-field, Landau level filling factor, and 
temperature dependence of the system's domain-wall soup and then
demonstrate  that this picture can account for many details
of the transport observations. 

Domain wall loops are characterized by their length and by their 
charge, with infinitely long loops appearing only for $b=0$ and 
$T > T_c$.  For finite size loops with a typical radius
larger than the domain wall width we can use our Hartree-Fock
self-consistent results for $\theta(X)$ to estimate the Coulomb
self-interaction energy. A two-dimensional charge density
of a circular loop with excess charge $e$ distributed
uniformly along the domain wall
is given by
\begin{equation}
\rho_{2D}(r)=\frac{e}{4\pi r}\frac{d}{dr}\cos\theta(r)\;.
\label{rho2d}
\end{equation}
Since the corresponding Coulomb self-interaction energy is
proportional to the square of the charge and approximately inversely proportional
to the length, charged domain wall loops have a higher energy and,
at integer filling factors, will always 
be less common than neutral domain wall loops.
Resistance spikes are generically observed slightly away 
from integer filling factors, however, and here the situation changes
because the domain wall loops can exchange charge with the rest of 
the 2D electron system.  
\begin{figure}[h]
\epsfxsize=3.3in
\hspace*{-0.5cm}
\centerline{
\epsffile{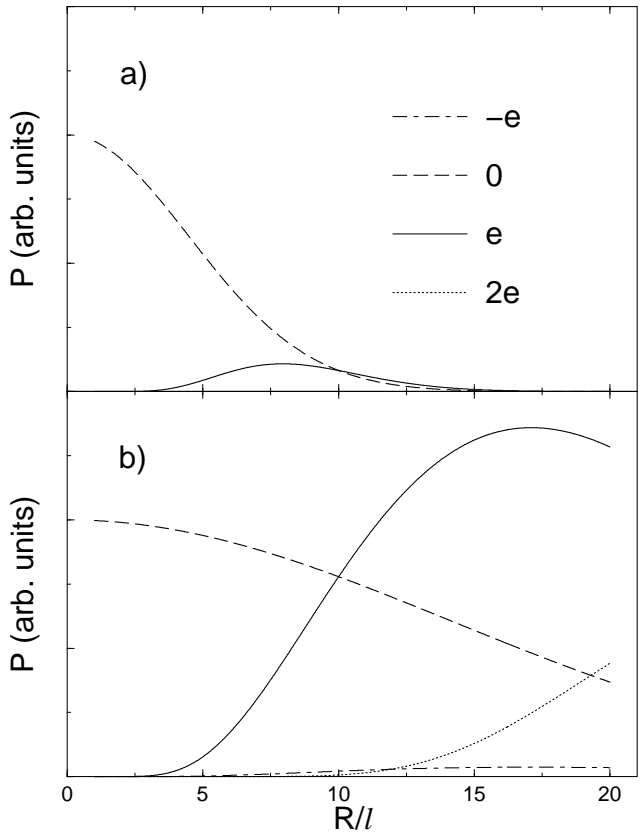}}


\caption{
Near--$T_c$, $\nu>3$,
statistical weights of domain wall loops with excess
charge from $-e$ to $2e$ are plotted as a function of the loop radius.
a) For $b/k_B\approx$~20mK, the effective external field  
corresponds to the edge
of the measured resistance spike. Small neutral  domain wall loops
dominate in this case.
b) For $b\approx 0$, the coincidence condition is approached 
corresponding to  the spike maximum. The typical domain
wall loops are large and carry an excess charge.
}
\label{prlfig2}
\end{figure}
The lowest energy elementary charged excitations of 
the $\nu=3$ ground state are ordinary
Hartree-Fock electron
and hole quasiparticles \cite{lilliehook},
not charged domain wall loops.
In systems with no disorder, the 
chemical potential lies in the middle of the Hartree-Fock gap when $\nu$ is an
integer  but moves quickly (by $\delta \mu$) toward the electron
quasiparticle energy for $\nu >3$ and toward the hole quasiparticle
energy for $\nu < 3$.  These chemical potential shifts will be reduced by
disorder which broadens the quasiparticle bands.
The change in chemical potential favors charge $Q$ 
over neutral skyrmions by a large factor $\exp (Q|\delta \mu||/k_B T)$.
This factor can be estimated quantitatively
using the experimental value of the quasiparticle excitation
gap ($\sim 2$~K) \cite{depoortere}.

Also important in controlling the domain wall soup is the 
effective field $b$, which measures the distance from Landau level
coincidence.  For $b \ne 0$ the  energy of a domain wall loop 
has a contribution proportional to $b$ and the number of condensate 
electrons contained within the loop:
\begin{equation}
E_b=-\frac{b}{2\ell^2}\int dr\, r\,[\cos\theta(r)-1]\; .
\label{eb}
\end{equation}
This contribution will decrease 
the number of large domain wall loops enclosing the minority phase
and
is independent of the charge carried by the loop. 
Summing up $E_b$, the Coulomb self-interaction and chemical potential
contributions and the Hartree-Fock domain wall energy of a circular loop
of radius $R$, $2\pi R\gamma$, we can estimate 
statistical weights of neutral and charged domain wall loops
in the sample of De Poortere {\em et al}.
In Fig.~\ref{prlfig2} we plot our results for temperature near 
$T_c$ and $\nu > 3$. For non-zero $b$-fields, small neutral domain wall
loops dominate, while typical loops at the coincidence are large and
carry an excess charge. 

\begin{figure}[h]
\epsfxsize=3.3in
\hspace*{-0.5cm}
\centerline{
\epsffile{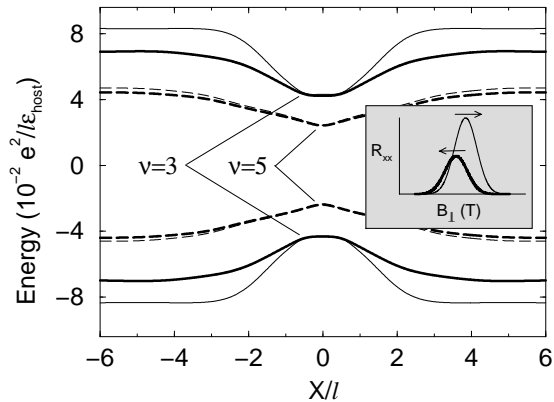}}

\vspace*{.7cm}

\caption{
Hartree-Fock quasiparticle energies, measured from 
the integer filling factor
chemical potential,
are plotted as a function of  orbit
guiding center for zero effective external field $b$.
For $|X|\gg0$ the spectrum describes ordinary, bulk electron and hole
quasiparticles. Near the domain wall center a
parabolicly dispersing 1D channel is formed. The thin and thick 
lines are obtained
assuming pseudospin-up and pseudospin-down majority 
polarizations, respectively.
At $\nu=3$ (solid lines),
the screening of the Coulomb interaction is substantially 
stronger in the latter
case resulting in different quasiparticle spectra for the two configurations.
The inset illustrates schematically the corresponding peak 
asymmetry in the measured
magnetoresistance near filling factor $\nu=3$. For comparison, we also show
in the main graph
spectra for $\nu=5$ (dashed lines).
The negligible difference between pseudospin-up
and pseudospin-down majority polarizations is consistent with the
similar peak heights recorded
in up and down magnetic field sweeps near $\nu=5$.
}
\label{prlfig3}
\end{figure}

We now address  characteristic features of the measured resistive hysteresis
loop. 
Dissipation can occur in Ising QHFs as a result
of Hartree-Fock quasiparticle diffusion, charged domain-wall-loop
diffusion, or as a result of charge diffusion within domain-wall
loops.   It is clear that the resistance spikes, which  
appear only for small $b$  and $T < T_c$, are associated with the 
appearance in the sample of large domain-wall loops.  Even though
these loops tend to be charged at the spike maximum, we expect 
that they will be immobile because of their large size and 
that dissipation due to their motion is small.
Instead we propose that mobile quasiparticles inside domain
walls are responsible for the increase of dissipation. In 
Fig.~\ref{prlfig3} we plot the
Hartree-Fock quasiparticle energies in a cross-section of a domain wall,
obtained from Eqs.~(\ref{hf})-(\ref{exchange}). In the
center of the domain wall the quasiparticle gap is reduced by nearly 50\%. 
Away from integer filling factors, the bottom of these 1D quasiparticle bands will 
lie below the chemical potential which is pinned to the bulk quasiparticle 
energies.  
We note that, unlike quantum Hall edge states, counter propagating
states exist within each loop.  
At $\nu\approx 3$, for example, 
the 1D states  have a nearly  parabolic dispersion characterized
by an effective mass $m^*\approx
2 m_e$. 
The particles  can 
cross the sample by scattering between overlapping loops. For $\nu=3$ and $T\approx
T_c$, it follows from inset of Fig.~\ref{prlfig1} and 
from Fig.~\ref{prlfig2} that the
characteristic loop radius at the spike edge is $3\ell$.
At low temperatures
or small magnetic lengths (high 2D electron gas densities), 
domain wall loops become  small and dilute, loops do not overlap,
and  charge diffusion
within domain walls cannot contribute to dissipation. 
This explains the absence of 
resistance spikes  at Landau level coincidence \cite{depoortere} under these
circumstances. 

The above mechanism also explains  the 
different resistance spike heights observed in up and down field
sweeps \cite{depoortere}
near $\nu=3$. 
As shown in the inset of Fig.~\ref{prlfig1},
the majority pseudospin Landau level for the up-sweep is the $n=2$ spin-up level while
for the down-sweep it is the $n=0$ spin-down Landau level. 
This difference in Landau level configurations in the two sweep directions
alters  remote Landau level screening in the sample
which has a marked effect on the quasiparticle energy spectrum.
The reduction of the quasiparticle gap in the domain wall, relative to its bulk value,
is stronger in the up-sweep case, leading to more domain-wall quasiparticles and 
more dissipation, as seen in experiment.
Similar agreement between the measured hysteresis loop properties
and domain wall quasiparticle spectra applies for 
$\nu=4$ \cite{depoortereunpubl}. 
In Fig.~\ref{prlfig3} we also plot energy spectra 
at $\nu=5$ which are nearly identical
for the up or down majority pseudospin orientations. 
This explains the absence of peak-height asymmetry in the hysteresis 
measurement 
at this filling factor $\nu=5$.\cite{depoortereunpubl}

We thank Etienne De Poortere, Herbert Fertig, and Mansour Shayegan for many
important discussions. Our work was supported by 
R.A. Welch
Foundation, by Minsistry of Education
of the Czech Republic under Grant OC P5.10, and by the
Grant Agency of the Czech Republic under Grant 202/01/0754.


\begin{references}
\bibitem{caveat} Self-consistent
local-spin-density approximation calculations were carried out to
establish that for the 15~nm wide AlAs quantum wells studied in these experiments,
orbital effects of the in-plane field are negligibly
small. In the calculation we assumed Land\'e g-factor $g=1.9$,
electron effective mass
$m^*=0.41 m_e$, and average heterostructure dielectric constant
$\epsilon = 11.5$.
\end{references}
\end{document}